# Report on Workshop III “Cited References Analysis Using CRExplorer” at the 18th International Conference of the International Society for Scientometrics and Informetrics (ISSI2021)

Robin Haunschild* and Lutz Bornmann*, **

* Max Planck Institute for Solid State Research

Heisenbergstraße 1,

70569 Stuttgart, Germany.

Email: R.Haunschild@fkf.mpg.de, L.Bornmann@fkf.mpg.de

** Administrative Headquarters of the Max Planck Society

Science Policy and Strategy Department

Hofgartenstr. 8,

80539 Munich, Germany.

Email: bornmann@gv.mpg.de

We have organized Workshop III entitled "Cited References Analysis Using CRExplorer" at ISSI2021. Here, we report and reflect on this workshop. The aim of this workshop was to bring beginners, practitioners, and experts in cited references analyses together. A mixture of presentations and an interactive part was intended to provide benefits for all kinds of scientometricians with an interest in cited references analyses.

Cited references analyses complement the traditional times cited analyses. Cited references analyses offer the possibility to focus the impact analysis on specific publication sets (e.g., research fields, topics, journals, or oeuvres of researchers). In contrast to the usual times cited analysis that measures citation impact on the complete bibliographic database, cited references analyses measure citation impact on the selected publication set only. A specific form of cited references analysis was proposed by Bornmann and Marx (2013). This new form of cited references analysis has been named reference publication year spectroscopy (RPYS, Marx, Bornmann, Barth, & Leydesdorff, 2014). One of the main areas of application of RPYS is the search for historical roots of research fields, topics, journals, or researchers. RPYS analyses are performed in different stages: In the first stage, the publication set of interest is collected with the references cited therein. In the second stage, the cited references are counted for every referenced publication year. In the third and final stage, the early referenced publication years with a rather large number of cited references are investigated. These "peak" years frequently point to single (or few) often referenced publications that can be interpreted as origins or historical roots of research fields. The program CRExplorer (see www.crexplorer.net) was introduced by Thor, Marx, Leydesdorff, and Bornmann (2016) for simplifying and supporting the latter two stages. Two years later, advanced indicators that provide new cited references analysis opportunities were included in the capabilities of CRExplorer (Thor, Bornmann, Marx, & Mutz, 2018).

Our workshop was structured as follows:

- In the first part, we provided an introduction into the topic of the workshop. As an example RPYS analysis, we analyzed the publications of Lutz Bornmann (n=324). We discussed the most pronounced peaks in the referenced publication years 1965, 1968, 2000, 2005, and 2008. Afterwards, Werner Marx told the story about how RPYS began in a pre-recorded video (see Figure 1). A more detailed version of this short story of RPYS can be found on Figshare (Marx, 2021).
- In the second part, two researchers presented RPYS analyses: (1) Peter Kokol (who unfortunately could not participate in the live session) contributed his study entitled "Identifying historical roots in paediatric echocardiography using RPYS" (Kokol, Zavrsnik, & Blazun Vosner, 2021) in a pre-recorded video. He presented empirical RPYS results and – most interestingly – a comparison of the results with the opinion of echocardiography experts as a validation approach. Although the experts were surprised by a few identified historical roots, they agreed upon reflection that those cited references are indeed important publications in the field. Some publications that were judged by the experts as important historical roots were not found by this RPYS study. A follow-up study is planned by Peter Kokol to resolve such differences. (2) Rüdiger Mutz presented his contribution entitled "How to identify different segments of the growth development of cited references statistically? The Higgs boson research as an example" (see also Barth, Marx, Bornmann, & Mutz, 2014). He identified five segments with different growth rates in the cited literature of Higgs boson research and suggested a segmented regression approach that could give some additional objective insights into the empirical structure of the sequence of cited reference counts. Within the five segments, he identified the historical roots (landmark papers that were very frequently referenced) of Higgs boson research.

- In the third and final part, we performed an interactive RPYS analysis on the papers published in the *Journal of Informetrics*. We explained the basic functionalities of CRExplorer as well as the more advanced features. The participants could ask their questions throughout the workshop that we answered and discussed.

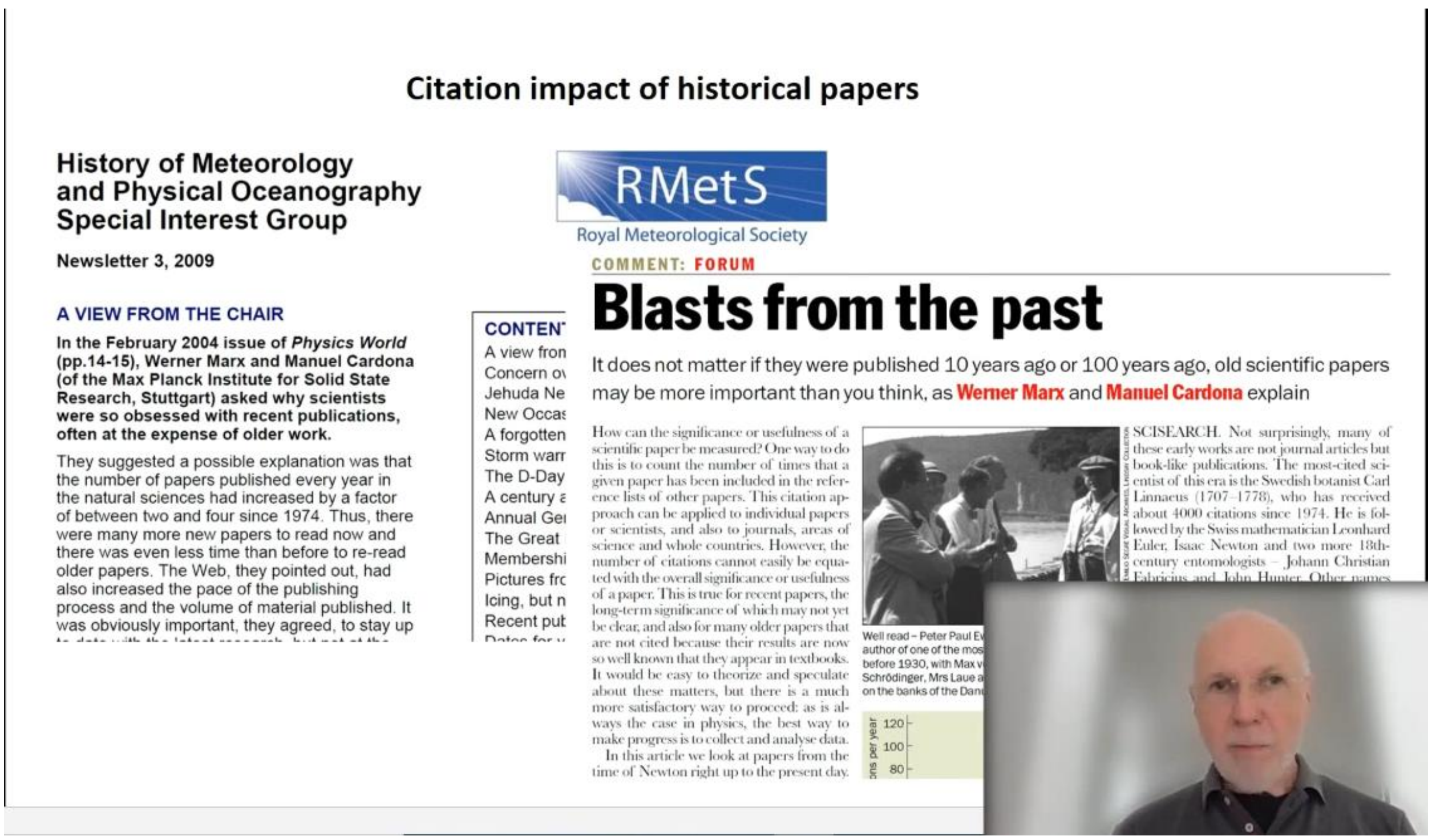

Figure 1: Screenshot of the pre-recorded video by Werner Marx telling the story of how RPYS evolved

We thank all participants and speakers for participating in and contributing to this workshop. Their interest in our workshop was indispensable for the success of our workshop. We hope that we were able to spark more interest in RPYS for future studies.